\documentclass[letter]{aa}
\usepackage{orcidlink}
\usepackage{graphicx}
\usepackage{txfonts}
\usepackage{amsmath,amssymb}
\usepackage{microtype}
\usepackage{hyperref}
\usepackage[nameinlink,capitalize]{cleveref}
\usepackage{csquotes}
\usepackage[switch]{lineno}

\newcommand{\Msun}{M_{\odot}}
\newcommand{\Mstar}{M_{\star}}

\newcommand{\xid}{\xi_{d}}
\newcommand{\xidzero}{\xi_{d,0}}
\newcommand{\epsstar}{\varepsilon_{\star}}
\newcommand{\Rc}{R_{\rm c}}
\newcommand{\fmech}{f_{\rm mech}}
\newcommand{\Reff}{R_{\rm e}}
\newcommand{\rs}{r_{\rm s}}
\newcommand{\Hg}{\mathcal{H}_{\rm gas}}
\newcommand{\Sg}{\Sigma_{\rm gas}}
\newcommand{\sg}{\sigma_{\rm gas}}

\newcommand{\dint}{d_{\rm int}}
\newcommand{\Rsb}{R_{\rm sb}}
\newcommand{\Msn}{M_{\star,{\rm SN}}}
\newcommand{\kabs}{\kappa_{\rm abs}}
\newcommand{\auv}{A_{{\rm UV},d}}
\newcommand{\pc}{\mathrm{pc}}
\newcommand{\kpc}{\mathrm{kpc}}
\newcommand{\kms}{\mathrm{km\,s^{-1}}}
\defcitealias{MartinezGonzalezetal2026}{Paper~I}

\begin{document}

\titlerunning{The first massive dusty galaxies}
\authorrunning{Mart\'\i nez-Gonz\'alez et al.}
\title{The first dusty galaxies across $6 \lesssim z \lesssim 14$}
\subtitle{Blue monsters, red monsters, and the bimodality of dust content in early galaxies}
\author{
  Sergio Mart\'inez-Gonz\'alez
  \thanks{Corresponding author. E-mail: sergiomtz@inaoep.mx}
  \inst{1}\orcidlink{0000-0002-4371-3823}
  \and
  Casiana Mu\~{n}oz-Tu\~{n}\'on
  \inst{2,3}\orcidlink{0000-0001-8876-4563}
  \and
  Santiago Jim\'enez
  \inst{4}\orcidlink{0000-0003-2808-3146}
}
\institute{
  Instituto Nacional de Astrof\'\i sica, \'Optica y Electr\'onica, AP 51, 72000 Puebla, M\'exico
\and
  Instituto de Astrof\'isica de Canarias, E 38200 La Laguna, Tenerife, Spain
\and
  Departamento de Astrof\'isica, Universidad de La Laguna, E 38205 La Laguna, Tenerife, Spain
\and
  Astronomical Institute of the Czech Academy of Sciences, Bo\v{c}n\'\i\ II 1401/1, 141 00 Praha 4, Czech Republic
}
\abstract
{JWST has revealed galaxies at $z>10$ with extremely blue UV-continuum slopes approaching $\beta_{\rm UV}\lesssim-2.4$, and low inferred dust contents. At similar redshifts, JWST and ALMA reveal dusty massive systems reaching $\beta_{\rm UV}\gtrsim-1$.}
{We aim to explain blue monsters and red monsters at $z\gtrsim10$ as different outcomes of supernova dust retention in clustered star-forming systems and connect this bimodality to dusty massive galaxies at $z\simeq6$--7.}
{We model dust removal in a  cluster-dominated regime through cloud-scale mechanical blowout followed by breakout from stratified high-redshift disks. The  shock-processed dust mass retained in the UV-attenuating layer is converted into a UV-continuum slope using absorption opacities derived from 3D hydrodynamical simulations of clustered supernovae in porous molecular clouds.}
{In compact, gas-rich galaxies, efficient cloud blowout and disk breakout reduce the retained dust fraction to blue-monster levels, yielding $\beta_{\rm UV}\lesssim-2.4$. Larger stellar masses increase the retained dust column and produce red-monster-like systems with $\beta_{\rm UV}\gtrsim-1.5$, reaching $\beta_{\rm UV}\simeq-0.5$ when radiative losses weaken large-scale driving.}
{The transition depends on whether clustered-supernova ejecta cross both the natal cloud and the galactic gas layer. Supernova dust production, shock processing, radiative losses, and mechanical venting can jointly explain UV-bright dust-poor galaxies and red dusty massive systems across $6\lesssim z\lesssim14$.}
\keywords{galaxies: high-redshift; galaxies: ISM; dust, extinction; galaxies: star formation; galaxies: evolution}
\maketitle
\section{Introduction}
\label{sec:intro}
JWST has allowed the identification of compact, UV-bright galaxies at $z>10$, with stellar masses of order $M_\star\simeq10^8$--$10^9\,\Msun$, extremely blue UV-continuum slopes, approaching $\beta_{\rm UV}\lesssim-2.4$, and dust-to-stellar mass ratios near or below $\log\xid\simeq -4$ \citep{ArrabalHaroetal2023,Ziparoetal2023,Ferraraetal2025}. These so-called blue monsters require efficient removal of recently produced dust from the UV-emitting regions. Recent models invoke  porosity of the interstellar medium, dust--star geometry, radiative feedback, grain growth, and low-opacity supernova dust to connect nearly attenuation-free UV-bright systems with dusty high-redshift galaxies \citep[and references therein]{Chevallardetal2013,Sommovigoetal2026,Burgarellaetal2026}. At similar epochs,  JWST/NIRSpec spectroscopy has confirmed EGS-z11-R0 at $z=11.452$,  the highest-redshift member of the emerging red monster population, with stellar mass $M_\star\simeq10^{9.2}$--$10^{9.6}\,\Msun$  and a red UV slope $\beta_{\rm UV}=-0.68\pm0.30$ \citep{Rodighieroetal2026}.  In the attenuation-free model, radiation pressure displaces dust from central star-forming regions, linking red and blue monsters through an obscured-to-cleared evolutionary sequence \citep{Ferraraetal2026}.  ALMA directly detects dust continuum in MACS0416-Y1 at $z=8.312$ \citep{Bakxetal2025}, while [O\,{\sc iii}] $88\,\mu{\rm m}$ detections in GHZ2 at $z=12.33$ and JADES-GS-z14-0 at $z=14.1793$ trace ionized gas and constrain the dust continuum through upper limits \citep{Zavalaetal2024,Schouwsetal2025}.
At cosmic dawn, the high gas densities and pressures in compact galaxies favor locally efficient star formation, with  the fraction of star formation in bound clusters, $\Gamma$, approaching unity \citep[e.g.][]{Vanzellaetal2026,Menonetal2026}. Clustered supernovae supply dust, then stellar winds and sequential shocks reshape its grain-size distribution, and collective superbubble flows transport it \citep{MartinezGonzalezetal2022}. 

\citet{MartinezGonzalezetal2026}, hereafter Paper I, showed that cloud-scale mechanical blowout removes much of this dust by combining 3D hydrodynamical survival yields with thin-shell blowout scalings \citep{Jimenezetal2019,Jimenezetal2021}. We now follow the mechanically vented dust beyond the parent cloud and test whether disk breakout carries it through the stratified gas layer into the halo.
Compact high-redshift disks provide large injection rates per unit area, while gas-rich systems can retain thick confining layers. We combine the disk-breakout thresholds of \citet{Royetal2013} with the cloud-scale retention model of \citetalias{MartinezGonzalezetal2026} and compute the UV slope from the retained, shock-processed dust.
\section{Model}
\label{sec:model}
We follow supernova dust through cloud-scale venting and disk breakout in a compact, gas-rich galaxy.  We then convert the dust retained within the UV-attenuating layer into a UV-continuum slope.
\begin{table}
\caption{Fiducial disk quantities for $M_{\star,{\rm gal}}=10^{8.5}\,\Msun$ and pre-breakout $f_{\rm gas}=0.80$.  Here sSFR and SFR are the specific and galaxy-wide star-formation rates; $\Msn$, $\Reff$, $\Sg$, and $\Hg$ are the stellar mass in supernova-phase clusters, effective radius, gas surface density, and vertical gas scale height. For $M_{\star,{\rm gal}}=10^{9.4}\,\Msun$, scale SFR and $\Msn$ by 8.3, $\Reff$ by 1.6, $\Sg$ by 3.2, and $\Hg$ by 0.31.}
\label{tab:scalings}
\centering
\small
\setlength{\tabcolsep}{3.5pt}
\begin{tabular}{@{}ccccccc@{}}
\hline\hline
$z$ &
sSFR &
SFR &
$\Msn$ &
$\Reff$ &
$\Sg$ &
$\Hg$ \\
&
Gyr$^{-1}$ &
$\Msun\,{\rm yr}^{-1}$ &
$\Msun$ &
pc &
$\Msun\,{\rm pc}^{-2}$ &
pc \\
\hline
6  & 4.10  & 1.30 & $3.89\times10^7$ & 655.1 & $1.32\times10^3$ & 315.1 \\
8  & 7.31  & 2.31 & $6.93\times10^7$ & 542.6 & $1.93\times10^3$ & 216.1 \\
10 & 11.59 & 3.67 & $1.10\times10^8$ & 466.8 & $2.60\times10^3$ & 159.9 \\
12 & 17.02 & 5.38 & $1.62\times10^8$ & 411.8 & $3.34\times10^3$ & 124.5 \\
14 & 23.66 & 7.48 & $2.24\times10^8$ & 369.9 & $4.14\times10^3$ & 100.4 \\
\hline
\end{tabular}
\end{table}
\begin{figure*}[t]
  \centering
  \includegraphics[width=0.9\textwidth]{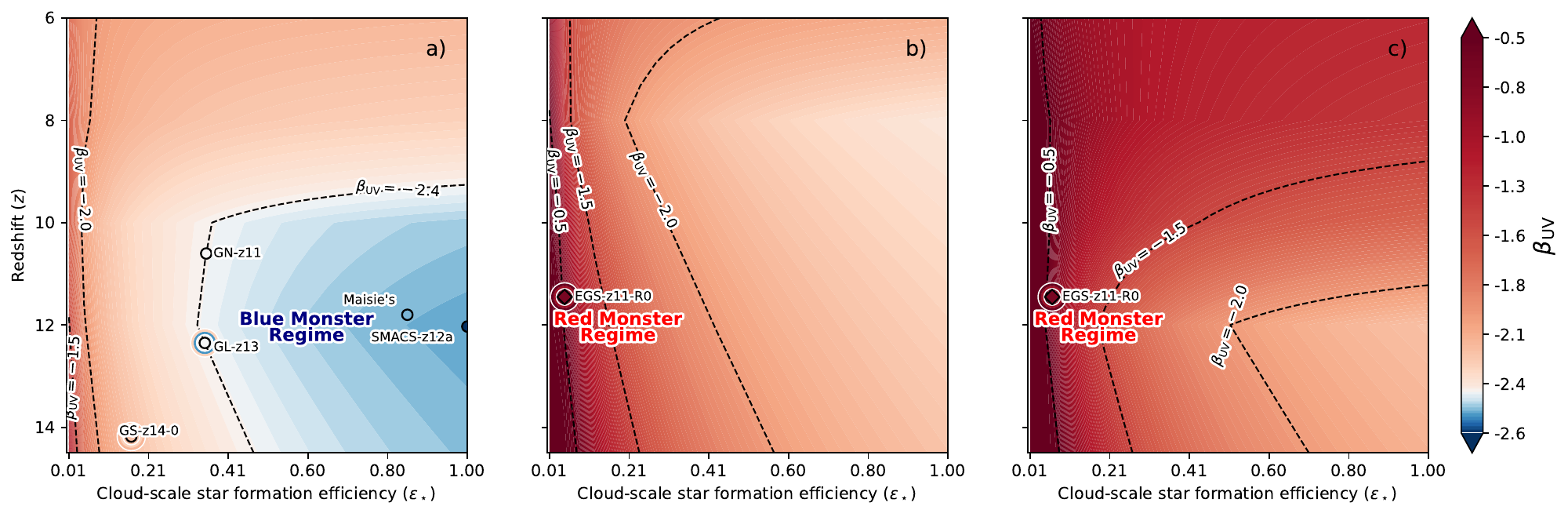}
\caption{UV-continuum slope in $(z,\epsstar)$ for $f_{\rm gas}=0.80$, $\sg=75\,\kms$, $t_{\rm SN}=30\,{\rm Myr}$, and $\Reff(z_{\rm ref}=2.25)=1.165\,\kpc$.  Here $\epsstar$, $\sg$, $t_{\rm SN}$, and $\Reff$ denote the cloud-scale star-formation efficiency, gas velocity dispersion, supernova-phase duration, and effective radius. Colors and dashed contours give $\beta_{\rm UV}$. Panel a uses $M_{\star,{\rm gal}}=10^{8.5}\,\Msun$; panels b and c use $10^{9.4}\,\Msun$, with 99 per cent radiative loss in panel c. Blue and red regions indicate efficient and inefficient dust venting. Markers show $\epsstar$ inferred from observed $(z,\beta_{\rm UV})$ values \citep{Ziparoetal2023,Bunkeretal2023,Finkelsteinetal2022,Carnianietal2024,Rodighieroetal2026}; colored rings show their $\beta_{\rm UV}$ uncertainties.}
\label{fig:beta_xid}
\end{figure*}
\subsection{Cloud-scale mechanical blowout and dust retention}
\label{subsec:retention}
We use the cluster-scale formalism of \citetalias{MartinezGonzalezetal2026}. A cluster of stellar mass $M_{\star}$ forms in a natal cloud of mass $M_{\rm gas,cl}=M_{\star}/\epsstar$ and cloud-core radius $\Rc$, where $\epsstar$ is the cloud-scale integrated star-formation efficiency.  The cloud solution defines the blowout radius, where the swept-up shell ceases to be pressure-confined, accelerates down the declining density gradient, and fragments, opening channels for gas and dust \citep{Jimenezetal2019,Jimenezetal2021}. Population averaging follows the truncated Schechter cluster mass function and Weibull cloud-core radius distribution adopted in \citetalias{MartinezGonzalezetal2026}. We adopt $\log\xidzero=-2.67$ for shock-processed supernova-condensed dust from the clustered-supernova simulations of \citet{MartinezGonzalezetal2022}, which include thermal sputtering, dust-induced cooling, shocked stellar winds in remnant cavities, and secondary shocks \citep{MartinezGonzalezetal2018}. Grain growth is followed but excluded from $\xidzero$ because seed grains are likely scarce at the highest redshifts.
For each stellar cluster, $\fmech(M_{\rm gas,cl},\Rc)$ is the fraction of supernovae exploding beyond the blowout radius. Cloud-scale mechanical blowout then leaves \begin{equation} {\xid}_{\rm cl}=\xidzero\left(1-\fmech\right), \label{eq:xid_cloud} \end{equation} where $\fmech$ is evaluated from the \citetalias{MartinezGonzalezetal2026} blowout scalings.
\subsection{Disk breakout and coalesced superbubbles}
\label{subsec:breakout}
After cloud-scale mechanical blowout, dust carried by a stellar-cluster superbubble reaches the halo through disk breakout only if the superbubble crosses the vertical scale height, $\Hg$, at supersonic speed \citep{Royetal2013}.  Disk breakout occurs when the swept-up shell reaches $\Hg$ before stalling; it then accelerates into the declining density profile, fragments, and opens a channel for hot gas \citep{TenorioTagleandMunozTunon1997,TenorioTagleandMunozTunon1998}; otherwise, the superbubble remains confined within the gas layer. Breakout leaves low-column holes \citep{Jimenezetal2024} and transports surviving grains out of the disk. For a stellar cluster of mass $M_{\rm SC}$, we assign a constant mechanical luminosity over the supernova phase, following \citet{Royetal2013},
\begin{equation}
L_{\rm SC}=6.3\times10^{35}
\eta_{\rm SN}M_{\rm SC}
\left(\frac{E_{\rm SN}}{10^{51}\,{\rm erg}}\right)
\left(\frac{t_{\rm SN}}{5\times10^{7}\,{\rm yr}}\right)^{-1}
{\rm erg\,s^{-1}} .
\label{eq:Lroy}
\end{equation}
Here $\eta_{\rm SN}=1/(95.5\,\Msun)$  is the number of core-collapse supernovae per unit stellar mass for a Kroupa initial mass function over $0.01$--$120\,\Msun$, with progenitors above $8\,\Msun$ \citep{Kroupa2001}. We adopt $t_{\rm SN}=30\,{\rm Myr}$ from low-metallicity massive-star tracks \citep{Szecsietal2022}. From a canonical $10^{51}\,{\rm erg}$ per supernova, $E_{\rm SN}=5\times10^{49}\,{\rm erg}$ drives the large-scale flow after $\sim95$ per cent radiative loss at the dense cluster-wind shell, where no Sedov--Taylor phase ever develops \citep{MartinezGonzalezetal2019,MartinezGonzalezetal2022}.
Disk breakout requires both the mechanical-luminosity and supernova-phase star-formation surface densities to exceed the thresholds of \citet{Royetal2013},
\begin{equation}
\small
\frac{L_{\rm SC}}{\pi\Hg^{2}}
\ge10^{-4}\,{\rm erg\,cm^{-2}\,s^{-1}},
\,\,\,\,
\frac{\Msn}{\pi\Reff^{2}t_{\rm SN}}
\ge 0.1\,\Msun\,{\rm yr^{-1}\,kpc^{-2}} .
\label{eq:breakout_thresholds}
\end{equation}
Here $\Msn$ is the total stellar mass in clusters undergoing the supernova phase, given by the galaxy stellar mass formed over $t_{\rm SN}$ at the specific star-formation rate (sSFR), and $\Reff$ is the effective radius (Sect.~\ref{subsec:zmodel}). 

Superbubbles may overlap before reaching $\Hg$. We estimate their interaction distance following \citet{Silichetal2002},
\begin{equation}
\Rsb=267\,{\rm pc}
\left(\frac{\langle L_{38}\rangle}{n_{0}}\right)^{1/5}
t_{{\rm SN},7}^{3/5},
\qquad
\dint=2\min(\Rsb,\Hg),
\label{eq:rint}
\end{equation}
where $\langle L_{38}\rangle=L_{\rm SC}(\langle M_{\rm SC}\rangle)/(10^{38}\,{\rm erg\,s^{-1}})$, $t_{{\rm SN},7}=t_{\rm SN}/10^{7}\,{\rm yr}$, and $n_0$ is the midplane gas density. Here $\langle M_{\rm SC}\rangle=1.57\times10^{5}\,\Msun$ is the mass-weighted mean of the cluster mass function adopted in \citetalias{MartinezGonzalezetal2026}, and $\Rsb$ is evaluated at $t=t_{\rm SN}$.
The central overlap parameter is
\begin{equation}
\eta_0=
\frac{\pi\dint^2\Msn}
{2\pi\rs^2\langle M_{\rm SC}\rangle},
\label{eq:eta0}
\end{equation}
where $\rs=\Reff/1.678$ is the exponential scale length.
Overlapping bubbles form a coalesced complex with the collective mechanical luminosity of their parent clusters. Disk breakout yields the channel-opening fractions, $f\sb{\rm break}^{\rm ind}$ for individual superbubbles and $f\sb{\rm break}^{\rm coll}$ for coalesced complexes.  Combining cloud-scale mechanical blowout with disk breakout gives the retained dust-to-stellar mass ratio for each case:
\begin{equation}
{\xid}\sb{x}=
\xidzero\left(1-\fmech f\sb{\rm break}^{x}\right),
\qquad
x\in\{{\rm ind},{\rm coll}\}.
\label{eq:xid_break}
\end{equation}
 Thus, the retained component comprises dust for which cloud-scale blowout fails and dust that undergoes cloud-scale blowout but subsequently fails to escape through disk breakout.
The  galaxy-averaged dust-to-stellar mass ratio retained within the UV-attenuating layer is then
\begin{equation}
\langle\xid\rangle\sb{\rm gal}=
\left(1-f\sb{\rm coll}\right)
\langle{\xid}\sb{\rm ind}\rangle
+
f\sb{\rm coll}
\langle{\xid}\sb{\rm coll}\rangle .
\label{eq:xid_mix}
\end{equation}
Here $f\sb{\rm coll}$ is the fraction of stellar clusters experiencing their supernova phase in coalesced regions,  and the brackets denote averages over the cluster mass and cloud-core radius distributions.  The cloud-scale blowout model yielded dust-to-stellar mass ratios consistent with the $\log\xid\sim-4$ values inferred for spectroscopically confirmed blue monsters \citepalias[Fig.~2 of][]{MartinezGonzalezetal2026}.
\subsection{Redshift-dependent disk structure}
\label{subsec:zmodel}
We evaluate the breakout quantities for a compact, gas-rich galaxy with $M_{\star,{\rm gal}}=10^{8.5}\,\Msun$, the geometric mean of the blue-monster mass range, and a pre-breakout gas fraction $f_{\rm gas}=0.80$,  chosen to represent the larger gas reservoir before venting; the $f_{\rm gas}\simeq0.68$ inferred for GL-z11 and GL-z13 provides a post-breakout reference \citep{Ziparoetal2023}. Following the 10-Myr high-redshift star-forming main sequence measured over $3\le z\le9$ by \citet{Simmondsetal2025}, we adopt
\begin{equation}
{\rm sSFR}=0.05
\left(\frac{M_{\star,{\rm gal}}}{10^{10}\,\Msun}\right)^{0.02}
(1+z)^{2.30}\,{\rm Gyr}^{-1}.
\label{eq:ssfr}
\end{equation}
The stellar mass in clusters transiting the supernova phase is $\Msn=f_{\rm SN}M_{\star,{\rm gal}}$, with $f_{\rm SN}=\min({\rm sSFR}\,t_{\rm SN},1)$.  We extrapolate this relation beyond $z=9$. Its extension to $z=14$ spans $\simeq250$ Myr, or $\simeq175$ Myr from $z=10$\footnote{Cosmic ages assume a flat $\Lambda$CDM cosmology with $H_0=67.66~{\rm km~s^{-1}~Mpc^{-1}}$, $\Omega_{\rm m}=0.3111$, and $\Omega_\Lambda=0.6889$ \citep{PlanckCollaboration2020}.}.
The effective radius is anchored at $z_{\rm ref}=2.25$ with $\Reff(z_{\rm ref})=1.165\,\kpc$, following the late-type normalization and redshift trend of \citet{vanDerWeletal2014}. We use
\begin{equation}
\Reff(z)=\Reff(z_{\rm ref})
\left(\frac{1+z}{1+z_{\rm ref}}\right)^{-0.75}.
\label{eq:resize}
\end{equation}
 We extrapolate this late-type relation beyond its $0<z<3$ calibration range. It agrees with $\Reff\propto(1+z)^{-0.71\pm0.19}$ inferred from JWST data to $z\sim8$ \citep{Ormerodetal2024} and gives $\Reff=0.37$--$0.66\,\kpc$ over $6\le z\le14$, consistent with sub-kpc sizes at $5<z<14$ \citep{Morishitaetal2024}.
 The galaxy gas mass is $M_{\rm gas,gal}=f_{\rm gas}M_{\star,{\rm gal}}/(1-f_{\rm gas})$. The gas surface density, vertical scale height, and midplane density are \begin{equation} \Sg(z)=\frac{M_{\rm gas,gal}}{2\pi\rs^{2}}, \,\, \Hg(z)=\frac{\sg^{2}}{\pi G\Sg}, \,\, n_0=\frac{\Sg}{2\mu_{\rm H}m_{\rm p}\Hg}. \label{eq:diskstruct} 
\end{equation}
Here $\mu_{\rm H}=14/11$, $m_{\rm p}$ is the proton mass, and $\sg=75\,\kms$ is the gas velocity dispersion adopted for vertical support, consistent with the $53$--$100\,\kms$ range inferred for luminous $z\simeq14$ galaxies \citep{Carnianietal2024}. The resulting disk properties are listed in Table~\ref{tab:scalings}.
 A disk configuration provides an idealized stratification length and breakout condition applicable to exponential gas layers and stratified spheroids. Irregularity and clumpiness should increase the scatter in dust retention: low-density channels favor earlier breakout; dense clumps delay expansion locally \citep{Royetal2013,Jimenezetal2019,Jimenezetal2021}.
\subsection{UV-continuum slope}
\label{subsec:uv}
 Because grain-size evolution can strongly affect the UV slopes of early galaxies \citep{Narayananetal2025}, we compute the absorption opacity, $\kabs(\lambda)$, from the grain-size distribution evolved through multiple sequential shocks in the clustered-supernova simulations of \citet{MartinezGonzalezetal2022}. We adopt an equal graphite-silicate mixture, $f_{\rm gra}=f_{\rm sil}=0.5$, using the graphite and astronomical-silicate optical constants of \citet{Draine1985}, with $\rho_{\rm gra}=2.26\,{\rm g\,cm^{-3}}$ and $\rho_{\rm sil}=3.3\,{\rm g\,cm^{-3}}$. The resulting mass-weighted opacity is $\kabs(1600\,\text{\AA})=1.45\times10^4\,{\rm cm^2\,g^{-1}}$. Over $1250$--$2600\,\text{\AA}$, it is well described by $\kabs(\lambda)\propto\lambda^{b_{\rm UV}}$, with $b_{\rm UV}=-0.48$,  and shows no pronounced $2175\,\text{\AA}$ bump. Both the UV-opacity normalization and spectral shape are direct predictions of the evolved, post-shock grain population.

 We identify the UV-attenuating region with the portion of the galactic gas layer projected over a UV-emitting area, $\auv=\pi(R_{{\rm UV},d})^2$, and adopt $R_{{\rm UV},d}=0.8\Reff$, motivated by simulated UV-to-stellar size ratios at  $6 \lesssim z \lesssim 10$ for $\Mstar\lesssim10^9\,\Msun$ \citep{Shenetal2024}. The attenuated continuum is
\begin{equation}
L_{\lambda,{\rm obs}}\propto
\left(\frac{\lambda}{1600\,\text{\AA}}\right)^{\beta_{\rm int}}
\exp[-\kabs(\lambda)\Sigma_{d,{\rm UV}}],
\label{eq:luvlambda}
\end{equation}
 where
$\Sigma_{d,{\rm UV}}=
\langle \xid \rangle_{\rm gal}M_{\star,{\rm gal}}/\auv$
is the mean dust surface density within the UV-attenuating region. We adopt $\beta_{\rm int}=-2.62$  as the intrinsic dust-free UV-continuum slope, consistent with the blue limit at $z>10.5$ \citep{Cullenetal2024}. We obtain $\beta_{\rm UV}$ by fitting $L_{\lambda,{\rm obs}}\propto\lambda^{\beta_{\rm UV}}$ in log-log space over the Calzetti continuum window, $1250$--$2600\,\text{\AA}$ \citep{Calzettietal1994}. 
At fixed gas mass, increasing gas compactness raises the density and radiative losses; if star formation contracts with the gas, one expects the clustered-supernova power to become more concentrated and the path to breakout to shorten, setting a bimodality between efficient venting and dust retention.
\section{Results}
\label{sec:results}
From $z=6$ to $14$, $\Hg$ falls from $315.1$ to $100.4\,\pc$. At $t_{\rm SN}$, $R_{\rm sb}$ typically exceeds $\Hg$, so neighbouring cluster-driven bubbles overlap before reaching this height. Over the same redshift range, the breakout fraction rises from $0.32$ to $0.57$ for individual superbubbles and from $0.72$ to unity for coalesced complexes. \Cref{fig:beta_xid} shows how a plethora of UV colors arises from clustered-supernova dust processing, with cloud-scale blowout setting the retained dust fraction and disk breakout the fraction vented toward the halo. Each panel averages over $256\times256$ cloud-scale realizations in $(M_\star,R_c)$ at each point in the $(z,\epsstar)$ plane. In panel a, systems with $M_{\star,{\rm gal}}=10^{8.5}\,\Msun$ enter the blue-monster regime at high redshift and high $\epsstar$, where $\beta_{\rm UV}\lesssim-2.4$ as dusty ejecta leave the UV-attenuating gas layer. At lower $\epsstar$, more supernova dust remains embedded in the disk, yielding $\beta_{\rm UV}\simeq-2$ to $-1.5$.
Panels b and c raise $M_{\star,{\rm gal}}$ to $10^{9.4}\,\Msun$, the geometric mean of the EGS-z11-R0 range \citep{Rodighieroetal2026}. At $z=11.452$, the adopted scaling gives ${\rm SFR}\simeq40.4\,\Msun\,{\rm yr^{-1}}$, consistent with the inferred $10$--$40\,\Msun\,{\rm yr^{-1}}$ range. Panel c further reduces the energy available for large-scale driving to $E_{\rm SN}=1.0\times10^{49}\,{\rm erg}$, corresponding to 99 per cent radiative loss from repeated supernova impacts on each cluster's wind-driven shell. The larger retained dust column produces a red-monster regime that extends over a wider range of $\epsstar$ when less supernova energy is available for large-scale driving. These systems reach $\beta_{\rm UV}\simeq-1.5$ to $-0.5$, encompassing the red-monster-like regime and EGS-z11-R0 \citep[$\beta_{\rm UV}=-0.68\pm0.30$;][]{Rodighieroetal2026}. 

 The resulting $\beta_{\rm UV}$ traces attenuation by the retained dust column along UV-emitting sightlines. Dust--star geometry and redistribution make attenuation degenerate with total dust content \citep{Sommovigoetal2025b}. The retained dust mass is $M_d=\langle\xid\rangle_{\rm gal}M_{\star,{\rm gal}}$. Far-infrared predictions require the distribution, opacity, temperature, and radiative transfer of dust beyond this layer; ALMA detectability thus depends jointly on dust content and extent \citep{Ferraraetal2025}.  The attenuation-free model uses super-Eddington radiation-driven outflows to move dust to large radii and derives $\beta_{\rm UV}$ from visual attenuation \citep{Ferraraetal2026}; our model follows cloud-scale blowout across the cluster population and breakout by individual and coalesced superbubbles, with $\beta_{\rm UV}$ set by the opacity of retained, shock-processed dust. Radiative and mechanical feedback can coexist, with radiation pre-clearing channels and enhancing mechanical venting \citep{MartinezGonzalezetal2026,Menonetal2026}.

In the attenuation-free model, EGS-z11-R0 and blue monsters form an obscured-to-cleared sequence, whereas their separation in our calculations is set by mechanical-venting efficiency. At cloud scale, a galaxy can host clusters spanning strong dust retention to efficient blowout; its integrated UV slope reflects their population-weighted balance, supporting coexistence of red- and blue-monster regimes. The Cosmic Gems galaxy illustrates that a cluster-dominated system with $\Gamma\simeq1$ can host radiative and mechanical feedback within the same brief evolutionary phase \citep{Vanzellaetal2026}, making it challenging to determine which mechanism dominates dust transport out of the UV-attenuating layer.
In all cases, toward $z\sim6$--7, larger gas scale heights and lower mechanical-power surface densities suppress disk breakout, allowing these systems to retain larger dust columns. Mechanically compressed supershells may then promote rapid grain growth in dense gas \citep{MartinezGonzalezetal2021}, contributing to the higher dust contents of ALMA-detected galaxies \citep{Inamietal2022,Sommovigoetal2022,Ferraraetal2022,Sommovigoetal2026}.
\section{Conclusions}
\label{sec:conclusions}
We developed a two-stage supernova dust-venting framework that unifies blue and red monsters within a  cluster-dominated mechanical picture ($\Gamma\simeq1$). Dust condenses and is processed inside natal clouds, after which clustered supernovae entrain the surviving grains in hot superbubble flows. UV colors are set by the dust remaining within the galactic gas layer after cloud-scale blowout and disk breakout.  The UV-opacity normalization follows directly from the grain-size distribution evolved through multiple sequential shocks. Fiducial $10^{8.5}\,\Msun$ systems reach the blue-monster regime when high $\epsstar$ and compact gas layers enable efficient dust removal, yielding $\log\langle\xid\rangle_{\rm gal}\lesssim-4$ and UV slopes near the dust-free blue limit. Systems with $10^{9.4}\,\Msun$ retain larger dust columns and produce $\beta_{\rm UV}\simeq-1.5$ to $-0.5$, comparable to red-monster-like systems.
This bimodality reflects mechanical retention. Efficient blowout and breakout remove dusty ejecta from the attenuating layer, producing blue UV continua. Larger stellar mass, lower cloud-scale star formation efficiency, stronger radiative losses, and turbulent support through a larger $\Hg$ retain more supernova-condensed dust in the disk, producing redder slopes.
At $z\simeq6$--7, grain growth in mechanically compressed dense gas may supplement the retained supernova dust, strengthening the connection with dusty massive ALMA galaxies.
\begin{acknowledgements}
The authors thank the anonymous referee for their constructive comments, which improved the manuscript. Sergio Mart\'\i nez-Gonz\'alez acknowledges support from the Fundaci\'on Occident and the Instituto de Astrof\'isica de Canarias under the Visiting Researcher Programme 2025 to 2028 agreed between both institutions. This work is part of the collaboration ESTALLIDOS, supported by the Spanish research grant PID2022-136598NB-C31 (ESTALLIDOS 8) from the Spanish Ministry of Science and Innovation. SJ acknowledges support by the Czech Ministry of Education, Youth and Sports, through the INTER-EXCELLENCE II program, project LUC24023, and by the institutional project RVO:67985815. The authors thankfully acknowledge the computer resources, technical expertise and support provided by the Laboratorio Nacional de Supercómputo del Sureste de México, SECIHTI member of the network of national laboratories.
\end{acknowledgements}
\vspace*{-3pt}
\bibliographystyle{aa} \bibliography{descendants}
\end{document}